\documentclass[a4paper]{jpconf}

\usepackage[T1]{fontenc}
\usepackage[utf8]{inputenc}
\usepackage[font=small,labelfont=bf]{caption}

\usepackage{graphicx}
\begin{document}
\title{Search for the Standard Model Higgs boson production in association with top quarks in $pp$~collisions at 8~TeV with the ATLAS detector}

\author{Mar\'ia Moreno Ll\'acer, on behalf of the ATLAS Collaboration}

\address{II. Physikalisches Institut, Georg-August-Universit\"at G\"ottingen, Germany}

\ead{maria.moreno.llacer@cern.ch}

\begin{abstract}
The search for the production of the Higgs boson associated with a pair of top quarks in the ATLAS experiment~\cite{ATLAS} is presented. It focuses on Higgs bosons decaying to $b\bar{b}$ and events containing two leptons (electrons and/or muons). It uses 20.3~fb$^{-1}$ of $pp$~collision data at $\sqrt{s}$~=~8~TeV collected with the ATLAS detector at the LHC in 2012. No significant excess of events is found and the 95\% CL observed (expected) limit is 7.0$\times$SM (4.3$\times$SM). After combining with the single lepton final state an observed (expected) limit of 4.1$\times$SM (2.6$\times$SM) with a best fit $\mu$ of 1.7$\pm$1.4 is obtained~\cite{ttHconfnote}.
\end{abstract}

\section{Introduction}

Since the discovery of the Higgs boson by the ATLAS and CMS collaborations in July 2012~\cite{HiggsATLAS,HiggsCMS}, the focus is placed on measuring its properties and establishing whether it is the scalar boson predicted by the Standard Model. The discovery was based on measurements of the decay of the Higgs to other bosons ($H\to\gamma\gamma$, $H\to$$ZZ$) and, since then, clear evidence has been established for $H\to~W^+W^-$ and the fermionic $H\to\tau^+\tau^-$ decay mode. Given that the bottom quark is the heaviest accessible fermion, the dominant decay mode is $H\to~b\bar{b}$.
However the overwhelming multijet background precludes the possibility of a search for Higgs bosons produced via gluon-gluon fusion or vector boson fusion (dominant production modes) in this channel. Instead, the analyses focus on associated production of Higgs boson with a vector boson or a pair of top quarks, which decay leptonically allowing for a clear signature to be used for triggering and background suppression. The later channel, $t\bar{t}H$($H\to~b\bar{b}$), also allows for the direct measurement of the top-Higgs Yukawa coupling as $\sigma_{t\bar{t}H}\propto~g^2_{t\bar{t}H}$.

\section{Analysis strategy}
This analysis studies the $t\bar{t}H$ production with $H\to~b\bar{b}$ and the dileptonic decay of the top quark pair.
The experimental signature determines the selection: exactly two leptons of opposite charge ($ee$, $\mu\mu$, $e\mu$) and at least two $b$-jets. The leading and subleading lepton must have $p_T>25$~GeV and $p_T>15$~GeV, respectively.
In the $e\mu$ category, the scalar sum of the transverse momentum of leptons and jets, $H_T$, is required to be above 130~GeV.
In the $ee$ and $\mu\mu$ event categories, the invariant mass of the two leptons, $m_{ll}$, is required to be larger than 60 (15)~GeV in events with exactly (more than) two $b$-tags to suppress contributions from the decay of hadronic resonances. A further cut is applied in these categories to reject events close to the $Z$~boson mass: $|m_{ll} - m_Z| > 8$~GeV.

The selected events are categorised according to the number of jets with $p_T>25$~~GeV ranging from two to at least four and the number of $b$-jets ranging from two to at least four.
A total of six independent regions are considered (see Figure~\ref{DataExp_Yields_Regions}).
They are analysed separately and combined statistically to maximise the overall sensitivity.
The three regions with the largest signal-to-background ratio (S/B) are (3j,~3b), ($\geq$4j,~3b) and ($\geq$4j,~$\geq$4b). These regions are referred to as ``signal-rich regions'', as they dominate the sensitivity to the signal, and they were blinded at the first stage of the analysis. The fraction of $H\to~b\bar{b}$ decays in the $t\bar{t}H$ signal in these regions is expected to be about 90\%. The remaining three regions considered are referred to as ``signal-depleted regions'' (low S/B) and consist of ($\geq$~2j,~2b). These are dominated by different backgrounds and are used to constrain systematic uncertainties, thus improving the background prediction in the signal-rich regions.

\section{Background modelling}
 
\begin{figure}[!h]
  \begin{center}
    \resizebox{0.95\textwidth}{!}{
      \includegraphics[width=0.3\textwidth]{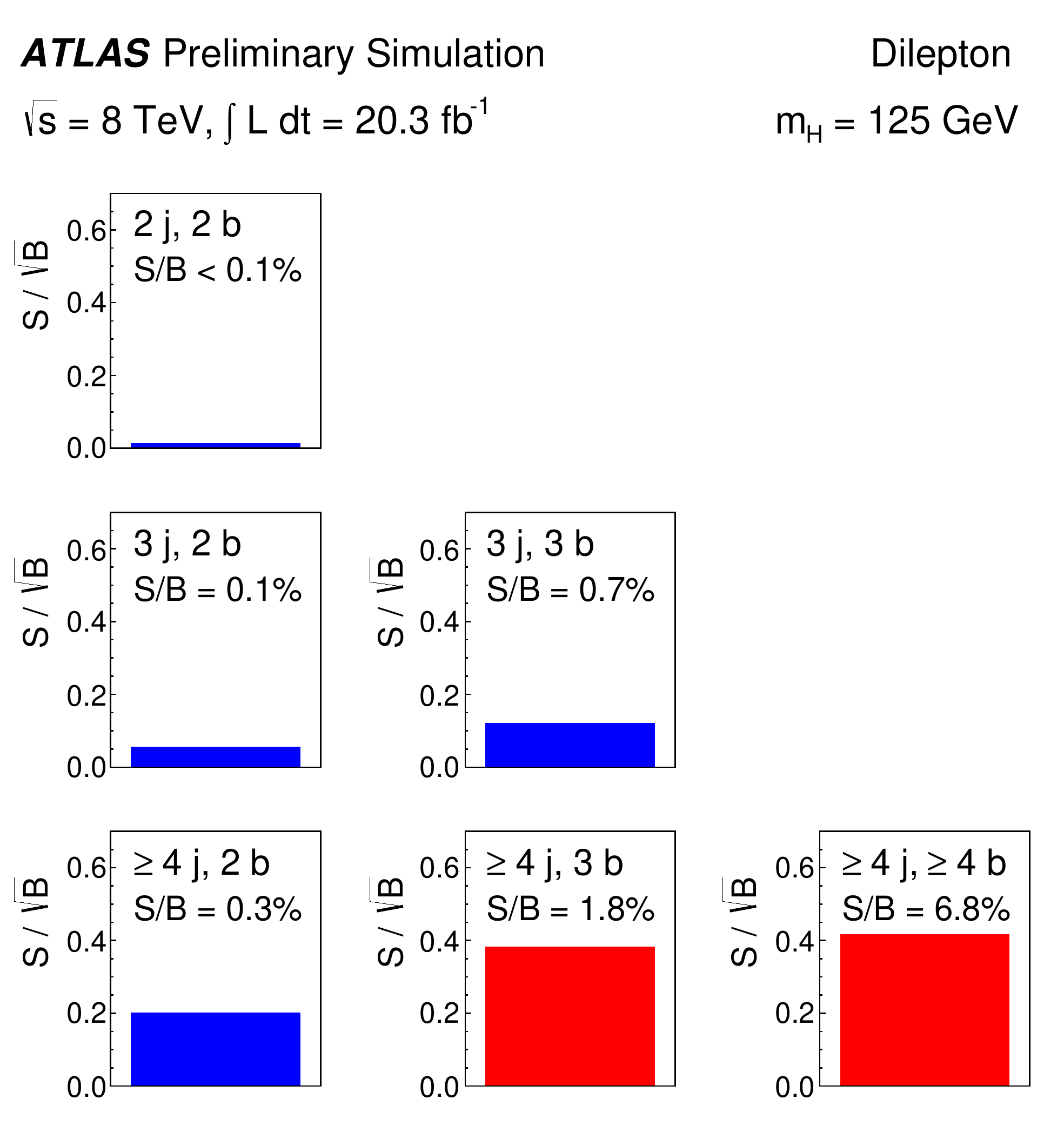}
      \includegraphics[width=0.38\textwidth]{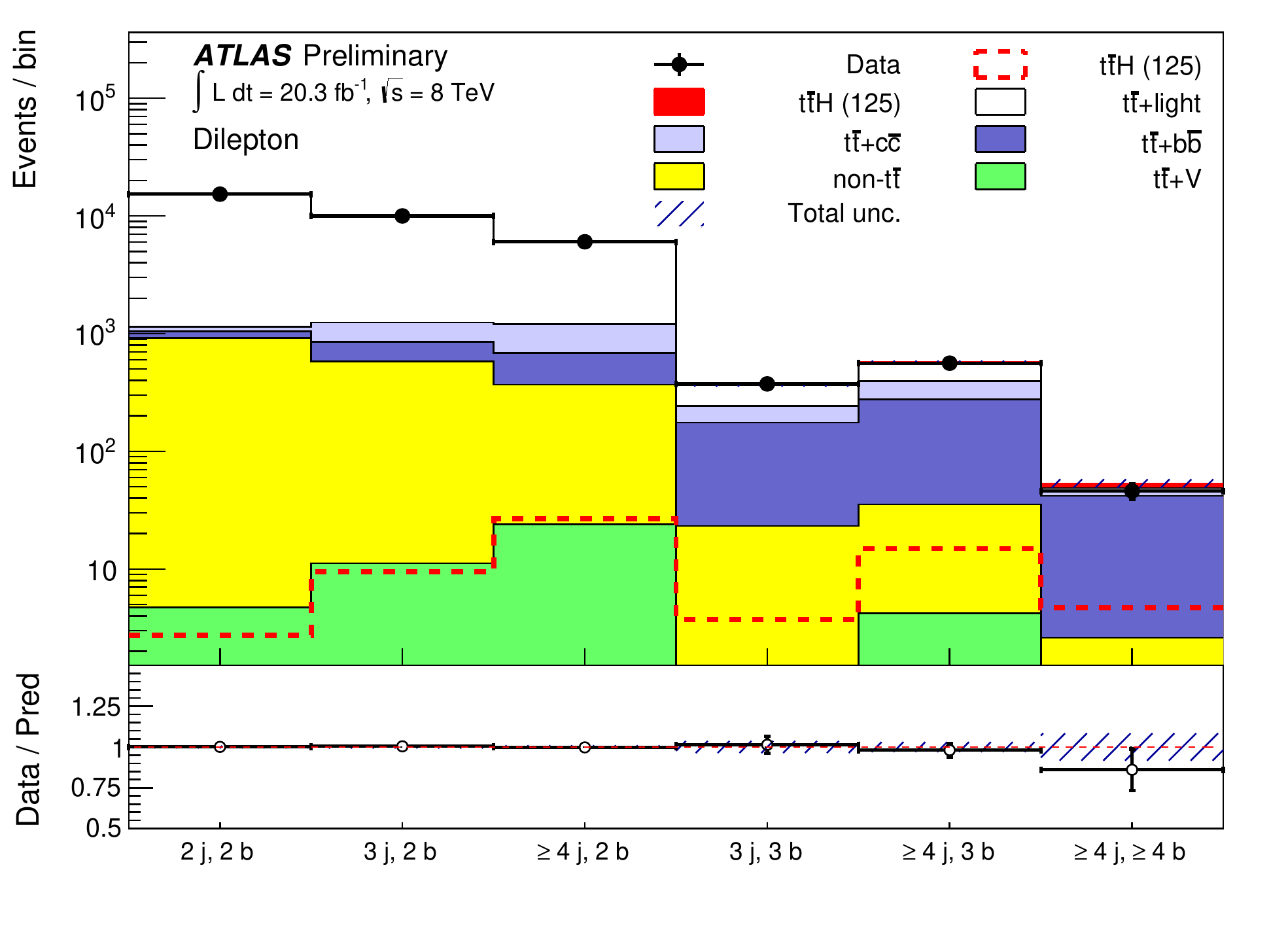}
    }
  \end{center}
 \caption{Left: The $S/\sqrt{B}$ and $S/B$ ratios for each of the analysis regions. Right: Comparison of prediction to data in all analysis regions after the fit to data in the dilepton channel. The signal, normalised to the SM prediction, is shown both as a filled red area stacked on the backgrounds and separately as a dashed red line. The hashed area corresponds to the total uncertainty on the yields. The non-$t\bar{t}$ backgrounds, represented in yellow, include the $W/Z$+jets, dibosons, single top and multijets.}
 \label{DataExp_Yields_Regions}
\end{figure}

As it can be seen in Figure~\ref{DataExp_Yields_Regions}, this analysis is largely dominated by the $t\bar{t}$~+~jets background. The $t\bar{t}$~+~light jets background dominates in the low $b$-tag multiplicity regions, while the higher $b$-tag multiplicity ones are dominated by $t\bar{t}$~+~$b\bar{b}$.
This background is modelled using Powheg+Pythia, where additional $b$ and $c$-quarks come from parton shower simulation.
The $t\bar{t}$~+~heavy-flavour modelling has been compared (both normalisation and kinematics) to the $t\bar{t}$~+~jets sample generated with MadGraph which includes the $t\bar{t}$~+~$b\bar{b}$ matrix elements.
To improve the agreement between data and simulation in $t\bar{t}$~+~jets dominated regions, events from Powheg+Pythia are corrected to reproduce the top quark $p_T$ and the $t\bar{t}$ system $p_T$ measured in data with correction factors based on the ratio of the measured differential cross sections at $\sqrt{s}$~=~7~TeV~\cite{TtbarDiffXS}.
Finally, the overall rate of $t\bar{t}$~+~$b\bar{b}$ and $t\bar{t}$~+~$c\bar{c}$ is calibrated to data using the background dominated bins.

\section{Signal-to-background discrimination}

\begin{figure}[!h]
  \begin{center}
    \resizebox{0.99\textwidth}{!}{
      \includegraphics[width=0.33\textwidth]{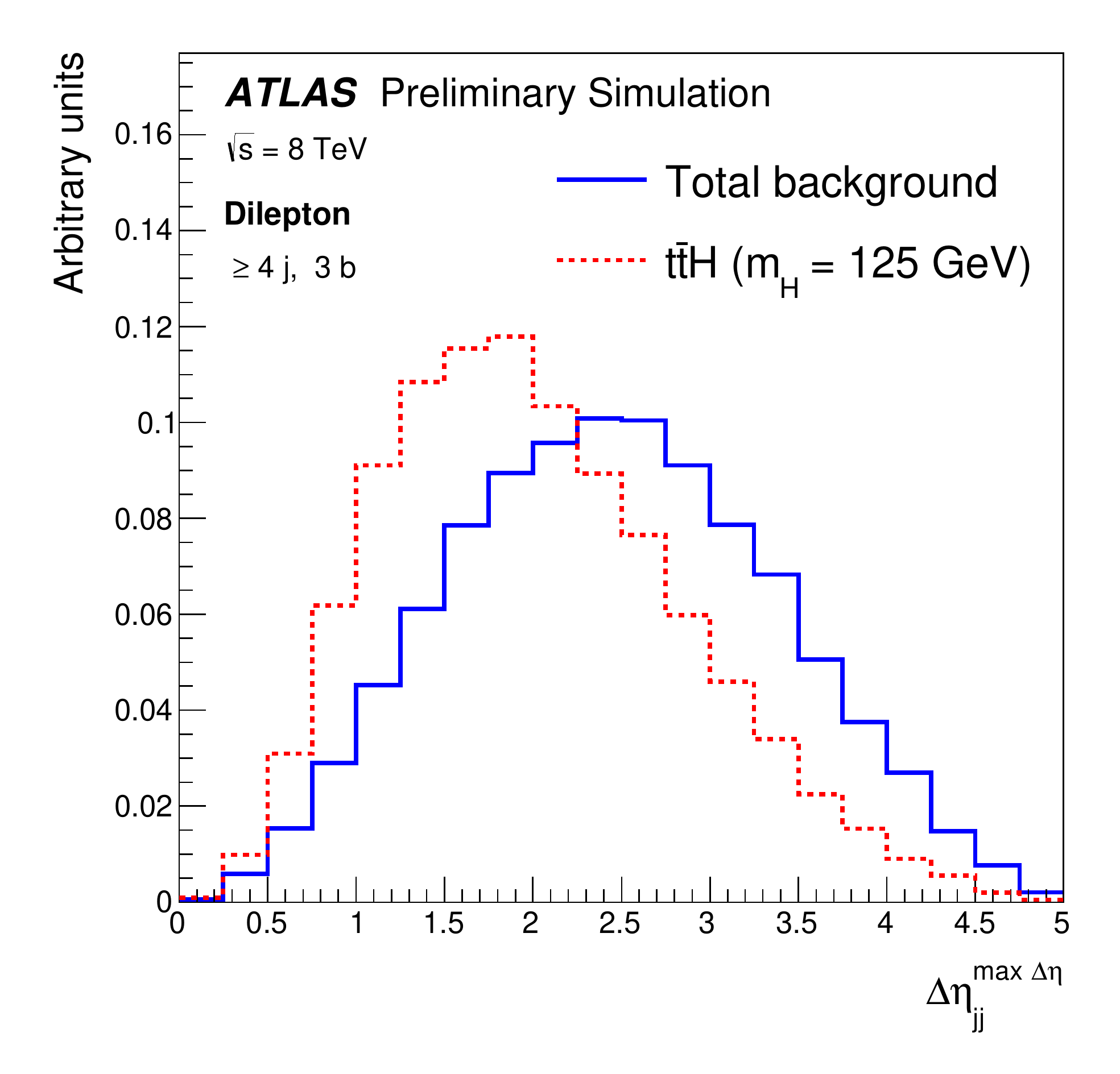}
      \includegraphics[width=0.33\textwidth]{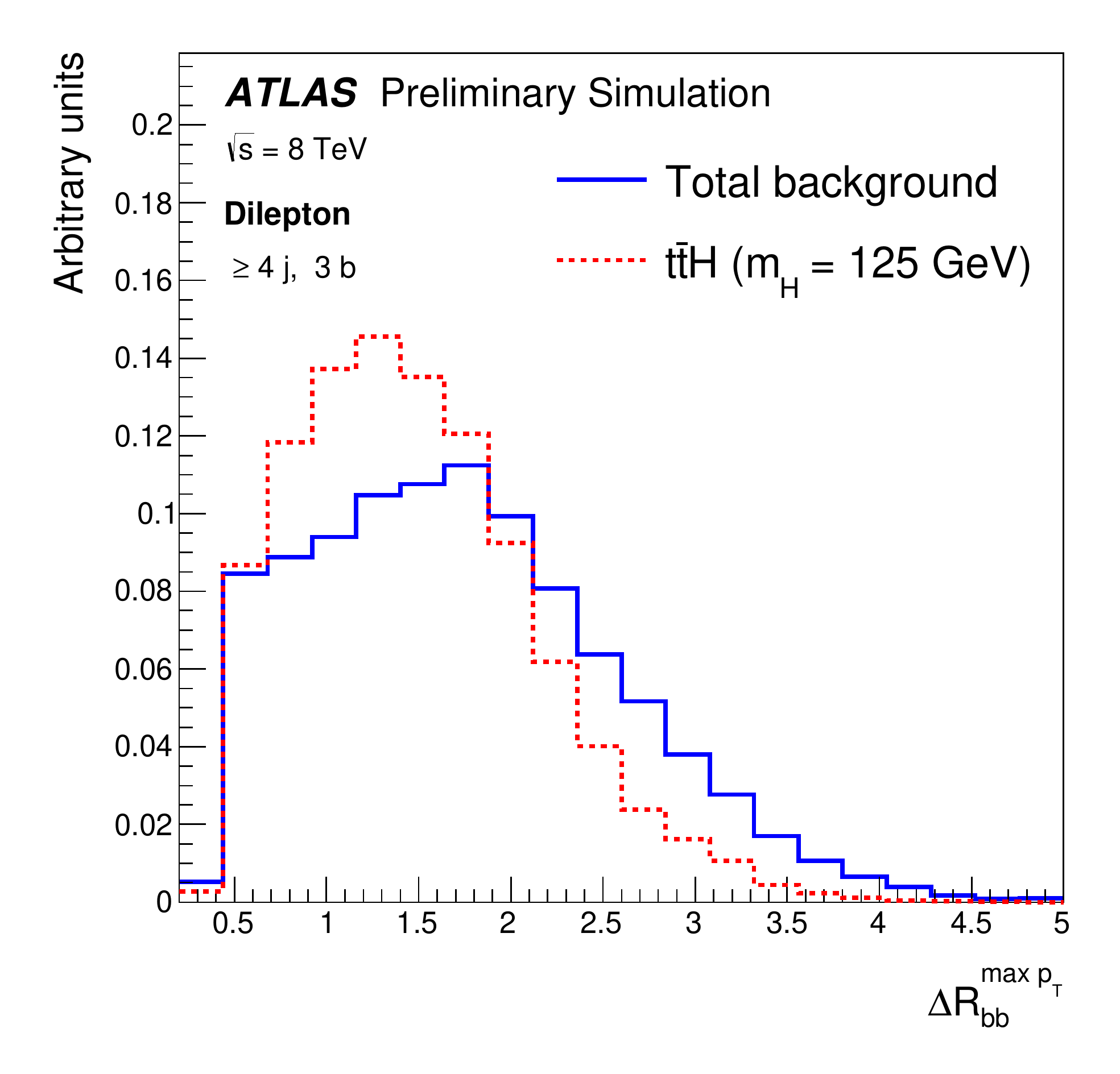}
      \includegraphics[width=0.33\textwidth]{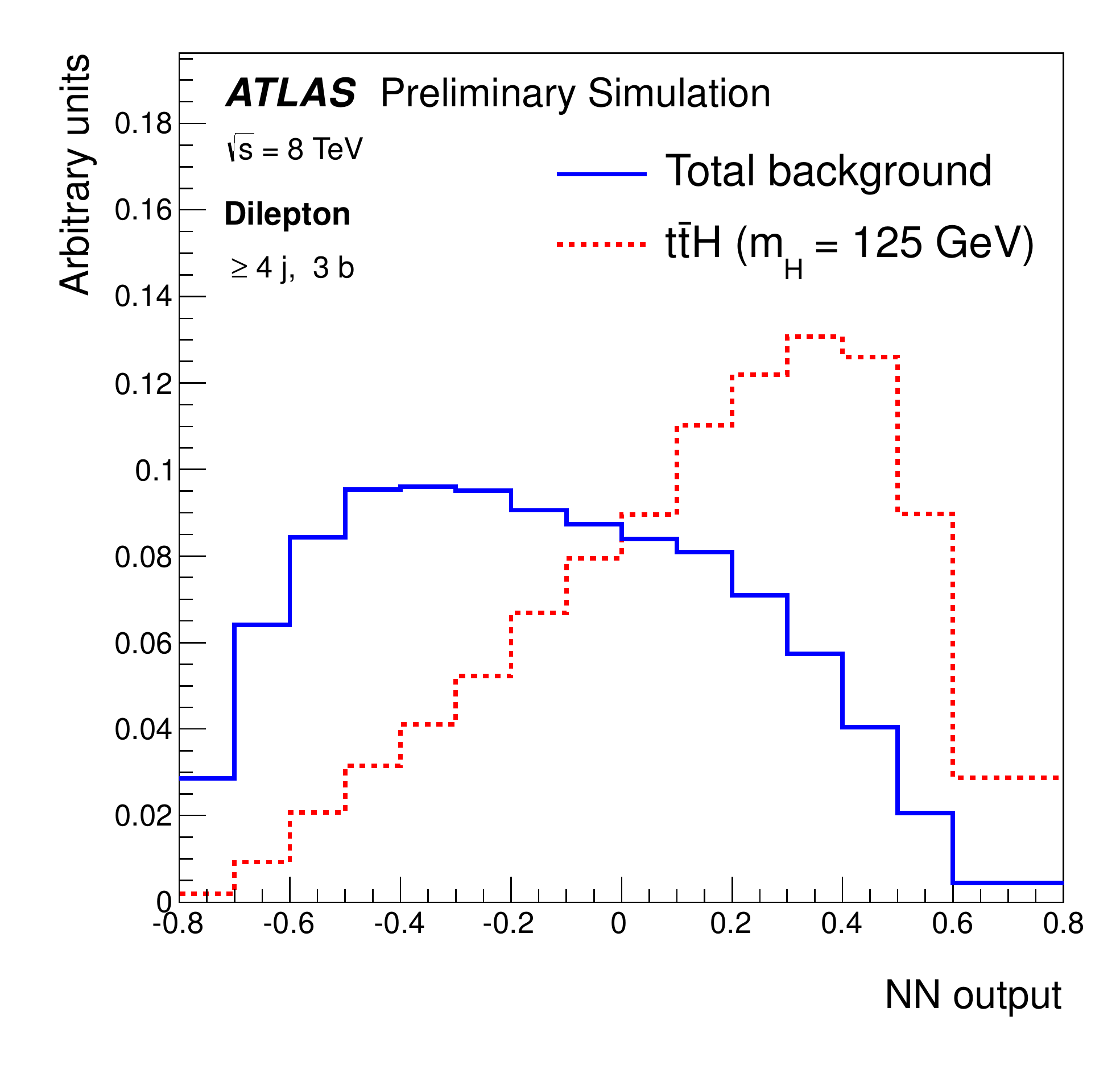}
    }
  \end{center}
  \caption{Signal (red, dotted) and background (blue, solid) separation plots for two of the NN input variables (maximum $\Delta\eta$ between two jets in the event and $\Delta$$R$ between the two $b$-jets with maximum momentum) and for the NN output in the ($\geq$4j,~3b) region. Distributions are normalised to unit area.}
  \label{VariablesInNN}
\end{figure}

In each analysis region, a suitable discriminating variable is chosen: a multivariate discriminant in the signal-rich regions and $H_T$ is taken in the signal-depleted ones.
In the signal-rich regions the output of a neural network (NN) implemented in NeuroBayes package~\cite{NNdoc} and trained to separate the signal, $t\bar{t}H$, from the main background, $t\bar{t}$+jets, is used.
Ten variables are selected for training the NN exploiting information involving single object or object pair kinematic properties, global event kinematics or event shape variables (see Figure~\ref{VariablesInNN}).


\section{Systematic uncertainties}

Several sources of systematic uncertainties are considered in this analysis, including instrumental and theoretical uncertainties.
These uncertainties can affect normalisation and/or shape of the signal and backgrounds. 
In order to reduce the degrading impact of systematic uncertainties on the sensitivity of the search, they are included in the fit via nuisance parameters.
For a given source, correlations between processes and across regions are taken into account.


The main instrumental systematic uncertainties arise from the jet energy scale and the flavour tagging efficiencies. 
The other major systematic uncertainties come from the modelling of the $t\bar{t}$+jets background: top $p_T$ and $t\bar{t}$ $p_T$ reweightings, the choice of the parton shower and hadronisation models and the normalisation uncertainty of $t\bar{t}$~+~heavy flavour.

\begin{figure}[!h]
  \begin{center}
    \resizebox{0.99\textwidth}{!}{
      \includegraphics[width=0.33\textwidth]{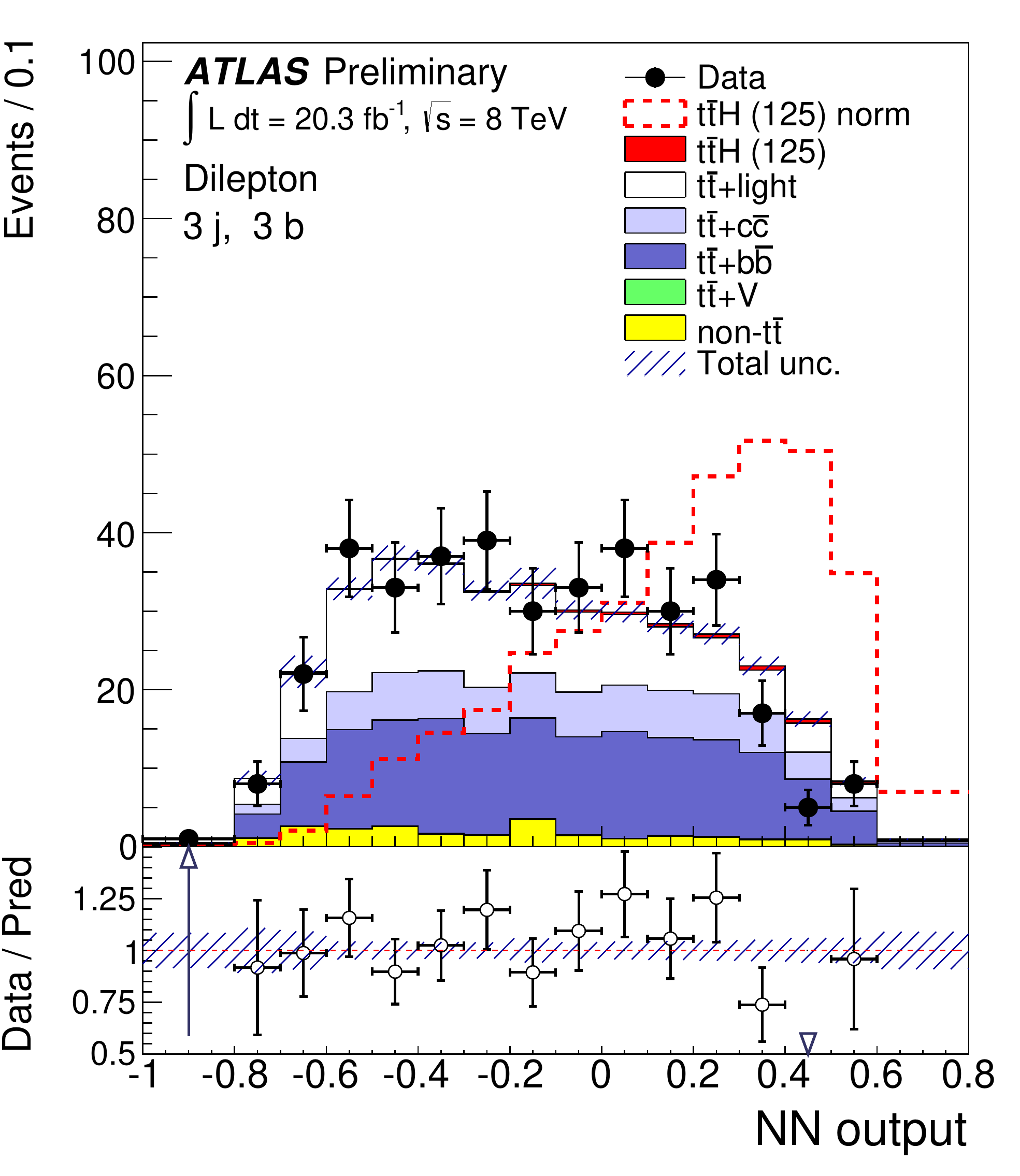}
      \includegraphics[width=0.33\textwidth]{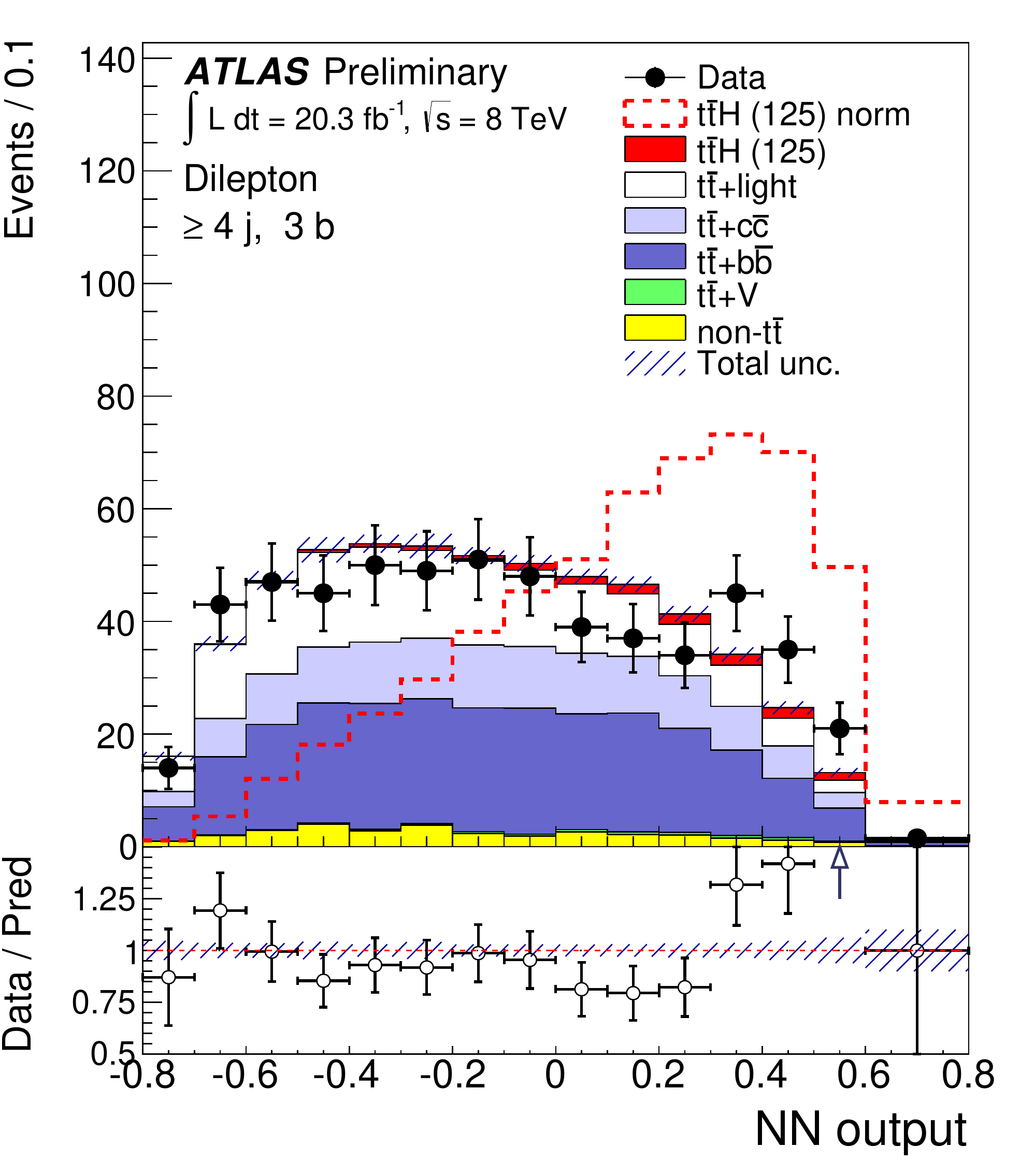}
      \includegraphics[width=0.33\textwidth]{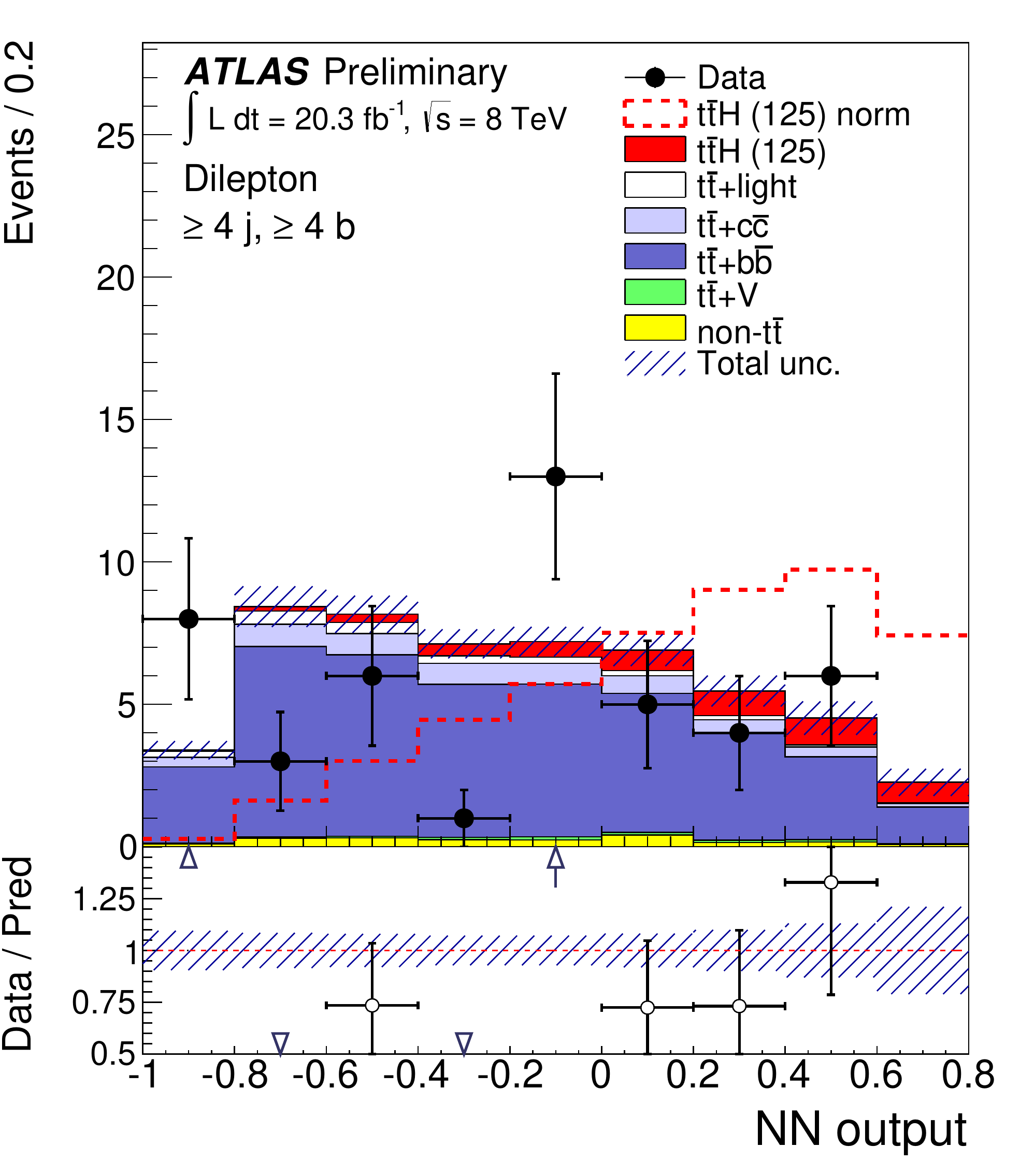}
    }
  \end{center}
  \caption{Final discriminant (NN output) for the most signal-rich regions: (3j,~3b), ($\geq$4j,~3b) and ($\geq$4j,~$\geq$4b). Background and signal are stacked in the filled histograms. Signal is superimposed normalised to background prediction. Data are represented by the black points.}
  \label{NNpostfit}
\end{figure}

\section{Statistical analysis}
\begin{figure}[!h]
  \begin{center}
    \resizebox{0.99\textwidth}{!}{
      \includegraphics[width=0.5\textwidth]{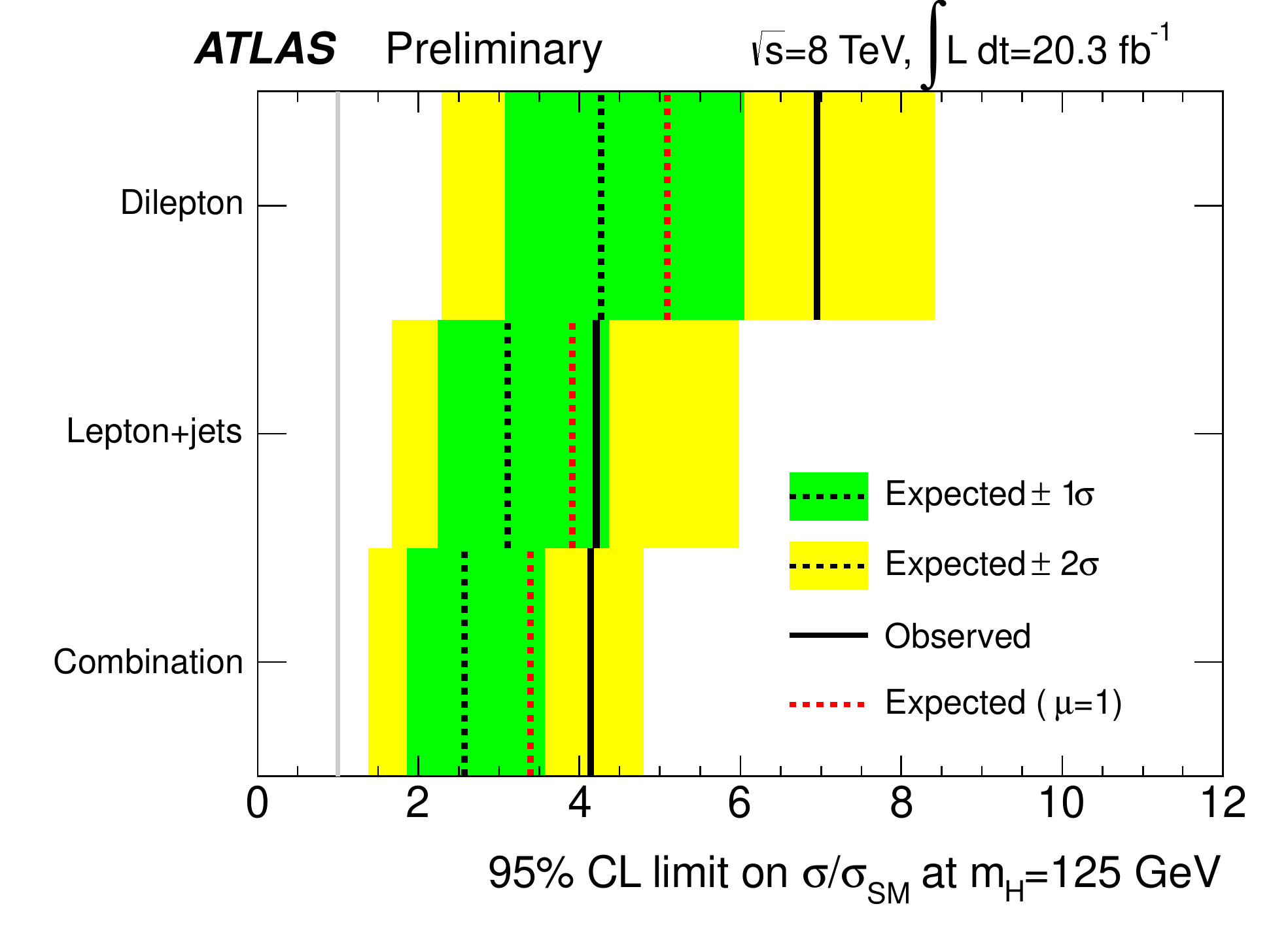}
      \includegraphics[width=0.5\textwidth]{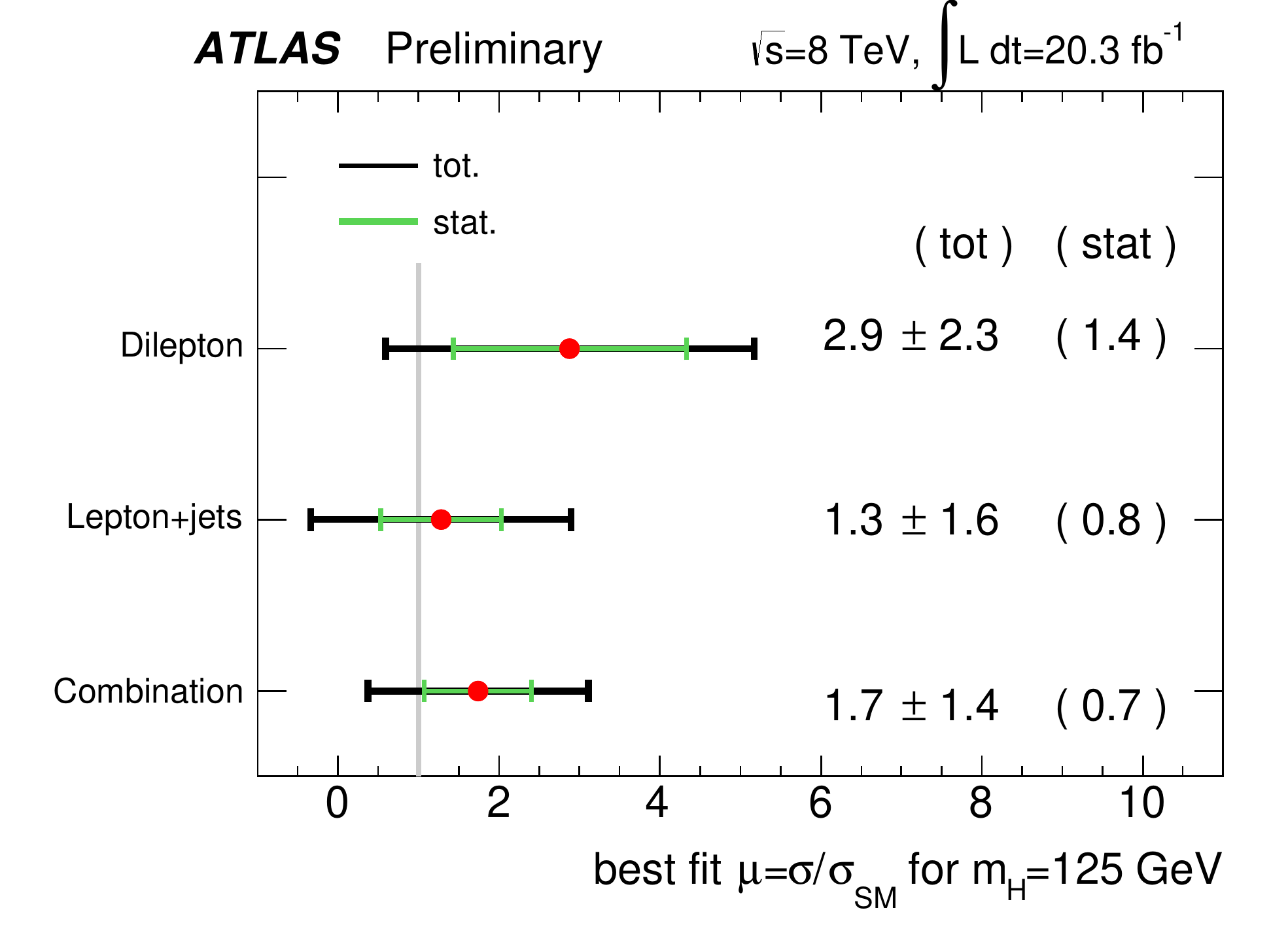}
    }
  \end{center}
  \caption{Left: 95\% CL upper limits for the dilepton and single lepton individual channels as well as their combination assuming $m_H$~=~125~GeV. The observed limits (solid line) are compared to the expected (median) limits under the background-only hypothesis (black dashed line) and under the signal-plus-background hypothesis assuming the SM prediction for $\sigma(t\bar{t}H)_{SM}$ (red dashed line). The surrounding bands correspond to the $\pm$1$\sigma$ and $\pm$2$\sigma$ ranges around the expected limit under the background-only hypothesis. Right: the observed signal strength and its uncertainty for the individual channels and the combination. Total uncertainties are shown in black while the statistical ones are represented in green.}
  \label{ResultsCombination}
\end{figure}

Figure~\ref{NNpostfit} shows a comparison of data and prediction for the discriminating variables for the signal-rich regions.
The distributions of the discriminants from each of the analysis regions considered are combined to test the presence of a signal assuming a Higgs boson mass of $m_H$~=~125~GeV.
The statistical analysis is based on a binned likelihood function $\cal{L}(\mu, \theta)$ constructed as a product of Poisson probability terms over all bins considered in the analysis, where $\mu~=~\sigma(t\bar{t}H)/\sigma(t\bar{t}H)_{SM}$ is the signal strength and $\theta$ is a set of nuisance parameters that encode the effect of systematic uncertainties.
A test statistic is used to measure the compatibility of the observed data with the background-only hypothesis (i.e. for $\mu$~=~0), and to make statistical inferences about $\mu$, such as upper limits, using the CLs method implemented in the RooFit package~\cite{RooFit}.
A simultaneous fit to the data is performed under the signal-plus-background hypotheses.
The fit improves the agreement between data and the prediction and reduces the background uncertainty by a factor 5 to 6 in the most sensitive regions.
No significant excess is found, and the 95\% CL observed (expected) limit is of 7.0 (4.3) times the SM prediction. After combining with the single lepton final state an observed (expected) limit of 4.1$\times$SM (2.6$\times$SM) with a best fit $\mu$ of 1.7$\pm$1.4 (see Figure~\ref{ResultsCombination}) is obtained. This result represents the most sensitive result obtained on data in the $t\bar{t}H$($H\to~b\bar{b}$) channel at the LHC.

\end{document}